\documentstyle[12pt,epsfig]{article}
\newcommand{\bfk}{\mbox{\boldmath $k$}}

\def\nostrocostruttino#1\over#2{\mathrel{\mathop{\kern 0pt \rlap
{\hbox{$#1$}}} \hbox{\kern-.135em $#2$}}}

\newcommand{\NP}[1]{{\it Nucl.\ Phys.}\ {\bf #1}}
\newcommand{\ZP}[1]{{\it Z.\ Phys.}\ {\bf #1}}
\newcommand{\PL}[1]{{\it Phys.\ Lett.}\ {\bf #1}}
\newcommand{\PR}[1]{{\it Phys.\ Rev.}\ {\bf #1}}

\newcommand{\beq}{\begin{equation}}
\newcommand{\eeq}{\end{equation}}
\newcommand{\barr}{\begin{eqnarray}}
\newcommand{\earr}{\end{eqnarray}}
\newcommand{\ba}{\begin{array}}
\newcommand{\ea}{\end{array}}
\def\lsim{\mathrel{\rlap{\lower4pt\hbox{\hskip1pt$\sim$}}\raise1pt\hbox{$<$}}}
\def\gsim{\mathrel{\rlap{\lower4pt\hbox{\hskip1pt$\sim$}}\raise1pt\hbox{$>$}}}

\newcommand{\qq}{q\bar q}

\newcommand{\la}{\lambda}
\begin{document}
\begin{flushright}
DFTT 16/98 \\
CPhT-S610-0498\\ 
INFNCA-TH9804 \\
hep-ph/9805234 \\
\end{flushright}
\vskip 1.5cm
\begin{center}
{\bf
Off-diagonal helicity density matrix elements for heavy vector mesons
inclusively produced in {\mbox{\boldmath $NN, \> \gamma N$}}
and {\mbox{\boldmath $\ell N$}} interactions.
}\\
\vskip 1.5cm
{\sf M. Anselmino$^1$, M. Bertini$^{1,2}$, F. Murgia$^3$ and
B. Pire$^{4}$}
\vskip 0.8cm
{$^1$Dipartimento di Fisica Teorica, Universit\`a di Torino and \\
      INFN, Sezione di Torino, Via P. Giuria 1, 10125 Torino, Italy\\
\vskip 0.5cm
$^2$Institut de Physique Nucl\'eaire de Lyon \\
43 Bvd. du 11 Novembre 1918, F-69622 Villeurbanne Cedex, France \\
\vskip 0.5cm
$^3$Dipartimento di Fisica, Universit\`a di Cagliari and \\ 
INFN, Sezione di Cagliari, CP 170, I-09042 Monserrato (Cagliari), Italy \\
\vskip 0.5cm
$^4$CPhT\footnote {CPhT is Unit\'e Mixte de Recherche C7644 du Centre 
National de la Recherche Scientifique}, Ecole Polytechnique, F-91128 Palaiseau,
France}\\

\end{center}
\vskip 1.5cm
\noindent
{\bf Abstract:}

\vspace{6pt}

\noindent
Final state interactions in quark fragmentation may give origin to non zero
values of the off-diagonal element $\rho_{1,-1}$ of the helicity density
matrix of vector mesons $V$ produced in current jets, with a large energy
fraction $x_{_E}$; the value of $\rho_{1,-1}(V)$ is related to the hard 
constituent dynamics and tests unusual properties of it. Some recent data on 
$\phi$, $K^*$ and $D^*$ produced in $e^+ e^-$ annihilations at LEP show such 
effects. Predictions are given here for $\rho_{1,-1}$ of heavy mesons produced
in nucleon-nucleon, $\gamma$-nucleon and $\ell$-nucleon interactions.
\newpage
\pagestyle{plain}
\setcounter{page}{1}
\noindent
{\bf 1 - Introduction}
\vskip 12pt
The elucidation of strong interactions requires a detailed understanding
of long distance properties of QCD. The difficulty of this physics
hinges on the non-perturba-tive nature of confinement phenomena. For
instance, whereas jet production is quite well under theoretical
control, the hadronization of quarks and gluons is still badly
understood.

In Ref. \cite{akp} it was suggested how the coherent fragmentation of
$\qq$  pairs created in $e^+ e^- \to \qq \to V + X$ processes might lead
to non zero  values of the off-diagonal element $\rho_{1,-1}$ of the
helicity density  matrix of the vector mesons $V$; in Ref. \cite{abmq}
actual predictions were  given for several spin 1 particles produced at
LEP energies in two jet events,  provided they carry a large fraction
$x_{_E}$ of the parent quark energy and have  a small intrinsic $\bfk_\perp$,
{\it i.e.} they are collinear with the parent jet.

The values of $\rho_{1,-1}(V)$ are related to the values of the off-diagonal
helicity density matrix element $\rho_{+-;-+}(\qq)$ of the $\qq$ pair,
generated in the $e^- e^+ \to \qq$ process \cite{abmq}:
\beq
\rho^{\,}_{1,-1}(V) \simeq [1 - \rho^{\,}_{0,0}(V)] \>
\rho_{+-;-+}(\qq) \label{old}
\eeq
where the value of the diagonal element $\rho^{\,}_{0,0}(V)$ can be taken from
data. The values of $\rho_{+-;-+}(\qq)$ depend on the elementary short distance
dynamics and can be computed in the Standard Model. Thus, a measurement
of $\rho_{1,-1}(V)$, is a further test of the constituent dynamics, more
significant than the usual measurement of cross-sections in that it depends
on the product of different elementary amplitudes, rather than on squared
moduli:
\beq
\rho^{\,}_{+-;-+}(\qq) = {1\over 4N_{\qq}} \sum_{\la^{\,}_{-}, \la^{\,}_{+}}
M^{\,}_{+-;\la^{\,}_{-} \la^{\,}_{+}} \> M^*_{-+; \la^{\,}_{-} \la^{\,}_{+}}
\,, \label{rhoz}
\eeq
where the $M$'s are the helicity amplitudes for the $e^- e^+ \to \qq$ process.
With unpolarized $e^+$ and $e^-$:
\beq
4N_{\qq} =
\sum_{\la^{\,}_q, \la^{\,}_{\bar q}; \la^{\,}_{-}, \la^{\,}_{+}} \vert
M^{\,}_{\la^{\,}_q \la^{\,}_{\bar q}; \la^{\,}_{-} \la^{\,}_{+}} \vert^2 \,.
\label{nqq}
\eeq
At LEP energy, $\sqrt s = M_{_Z}$, one has \cite{abmq}
\beq
\rho^{\,}_{+-;-+}(\qq) \simeq
\rho^{Z}_{+-;-+}(\qq) \simeq {1\over 2} \> {(g^2_{_V} - g^2_{_A})_q \over
(g^2_{_V} + g^2_{_A})_q} \, {\sin^2\theta \over 1+ \cos^2\theta} \, \cdot
\label{rhozap}
\eeq

Eq. (\ref{old}) is in good agreement with OPAL Collaboration data on
$\phi$, $D^*$ and $K^*$, including the $\theta$ dependence induced by Eq.
(\ref{rhozap}) \cite{opal1, opal2}; however, no sizeable value of
$\rho_{1,-1}(V)$ for $V= \rho, \phi$ and $K^*$ was observed by DELPHI
Collaboration \cite{delphi}. Further tests are then necessary.

We consider here other interactions in which the value of $\rho_{1,-1}(V)$
could be measured, namely $NN \to VX$, $\gamma N \to VX$ and $\ell N \to
\ell VX$,  with $V= \phi, D^*$ or $B^*$. The choice of a heavy vector
meson implies the dominance in each of these cases of a particular
elementary hard contribution, respectively
$gg \to \qq$,
$\gamma g \to \qq$ and $\gamma^* g \to \qq$, with $q = s, c $ or
$b$. The hadronization process -- the fragmentation of a $\qq$
pair -- is then similar to the one occurring in $e^+e^-$ annihilations;
however, the value of $\rho_{1,-1}(V)$ in these cases should be different
from that observed in $e^-e^+ \to VX$ at LEP, due to a different underlying
elementary dynamics, {\it i.e.} a different value of $\rho^{\,}_{+-;-+}(\qq)$.
A measurement of $\rho_{1,-1}(V)$ in agreement with our predictions in these
other processes would be an unambigous test of both the quark hadronization
mechanism and the real nature of the constituent interactions.

\vskip 12pt
\goodbreak
\noindent
{\bf 2 -} {\mbox{\boldmath $\rho^{\,}_{1,-1}(V)$}} {\bf in the process}
{\mbox{\boldmath $NN \to V + X$}}
\nobreak
\vskip 12pt
According to the QCD hard scattering scheme and the factorization theorem
\cite{colnoi, old} the helicity density matrix of a heavy hadron $V$
inclusively produced at large $p_T$ in the $AB \to VX$ process, with
unpolarized initial particles, at  leading twist and lowest order in the
coupling constant, can be written as
\barr
\rho^{\,}_{\la^{\,}_V,\la^\prime_V}(V) &\>& \!\!\!\!\!\!\!\!\!
d\sigma^{AB \to V X} =
\sum_{a,b,c,d,X} \> \sum_{\la^{\,}_c, \la^{\,}_d, \la^{\prime}_c,
\la^{\prime}_d, \la^{\,}_X} \label{gen} \\
& & f_{a/A}(x_1) \> f_{b/B}(x_2) \> d\hat\sigma^{ab \to cd}
\> \rho^{\,}_{\la^{\,}_c \la^{\,}_d; \la^\prime_c \la^\prime_d}(cd) \>
D^{\,}_{\la^{\,}_V \la^{\,}_X; \la^{\,}_c \la^{\,}_d} \>
D^{*}_{\la^{\prime}_V \la^{\,}_X; \la^\prime_c \la^\prime_d} \nonumber
\earr
where the $f_{i/I}$ are the unpolarized quark or gluon number densities,
$d\hat\sigma$ is the unpolarized cross-section for the elementary
interaction $ab \to cd$, and $\rho(cd)$ is the helicity density matrix of 
the $cd$ final partons.
Finally, $D^{\,}_{\la^{\,}_V \la^{\,}_X; \la^{\,}_c \la^{\,}_d}$ is the
helicity amplitude for the fragmentation process $cd \to VX$;
\beq
D_{cd}^{V,\la^{\,}_V} = \sum_{\la^{\,}_c,\la^{\,}_d,
\la^{\prime}_c,\la^{\prime}_d; X, \la^{\,}_X}
\rho^{\,}_{\la^{\,}_c \la^{\,}_d; \la^\prime_c \la^\prime_d} \>
D^{\,}_{\la^{\,}_V \la^{\,}_X; \la^{\,}_c \la^{\,}_d} \>
D^{*}_{\la^{\,}_V \la^{\,}_X; \la^{\prime}_c \la^{\prime}_d}
\label{polfra}
\eeq
is the fragmentation function of the $cd$ state, with polarization
described by $\rho(cd)$, into the hadron $V$ with helicity $\la^{\,}_V$.
If one chooses hadron $V$ collinear with, say, $c$, one has the usual
fragmentation function
\beq
D_c^V(x_{_E}) = \sum_{\la^{\,}_V} D_{c,d}^{V,\la^{\,}_V} \,,
\label{usfra}
\eeq
where $x_{_E}$ is the energy or momentum fraction
of parton $c$ carried by the observed hadron $V$.

Notice that we have considered the fragmentation process as $cd \to VX$,
neglecting all possible interactions of the fragmenting quarks or gluons
with the remnants of hadrons $A$ and $B$; we trust this to be a good
approximation for all events in which two well defined large $p_T$ jets,
originated by $c$ and $d$, are observed. Our considerations will be
limited to these cases.

Eq. (\ref{gen}) can be evaluated in any frame; we consider it in the
{\it c.m.} frame of interacting partons $a$ and $b$, which can be determined
provided one measures the two current jets total energy $\sqrt{\hat s}$ and
longitudinal momentum $p_L$
\beq
x_1 x_2 = {\hat s \over s} \quad\quad\quad\quad
x_1 - x_2 = {2p_L \over \sqrt s} \,\cdot
\label{newf}
\eeq
Our estimate for the helicity density matrix of $V$ will hold in such a
frame; the two body decays of $V$, which allow to measure $\rho_{0,0}(V)$
and $\rho_{1,-1}(V)$, should be observed in the helicity rest frame
\cite{bls} of $V$, as reached from the {\it c.m.} frame of $a$ and $b$.

In this frame the fragmentation process $cd \to VX$ is a {\it c.m.} forward
process, similarly to what happens in $e^-e^+ \to \qq \to VX$ \cite{abmq}.
We can then use the same considerations introduced in Ref. \cite{abmq} about
the otherwise unknown hadronization mechanism: by selecting $V$ collinear
to the parent jet, angular momentum conservation requires for the
forward fragmentation amplitudes appearing in Eq. (\ref{gen}):
\beq
\la^{\,}_X = \la^{\,}_V - (\la^{\,}_c - \la^{\,}_d) =
\la^{\prime}_V - (\la^{\prime}_c - \la^{\prime}_d)\,.
\label{las}
\eeq
All other helicity configurations are suppressed by powers of
$k_\perp/(x_{_E}\sqrt s)$, where $\bfk_\perp$ is the transverse momentum
of the observed hadron with respect to the parent jet. By choosing
vector mesons with a sizeable fraction of the quark energy, say
$x_{_E} \gsim 0.5$ and a $k_\perp$ such that $k_\perp/(x_{_E}\sqrt s) \ll 1$
we can safely neglect contributions from helicity fragmentation
amplitudes which do not satisfy Eq. (\ref{las}).

Let us now consider in particular the production of a vector meson, with a 
large $x_{_E}$ value and a small $k_\perp$, for which the corresponding 
dominant hard process is $gg \to \qq$; this should apply to the production of 
$\phi$ ($q=s,\bar s$), $D^*$ ($q=c,\bar c$) and $B^*$ ($q=b,\bar b$).

For the fragmentation process of a $\qq$ state Eq. (\ref{las}) implies
\beq
\sum_{\la^{\,}_q, \la^{\,}_{\bar q}, \la^{\prime}_q, \la^{\prime}_{\bar q},
\la^{\,}_X}
\rho^{\,}_{\la^{\,}_q \la^{\,}_{\bar q}; \la^\prime_q \la^\prime_{\bar q}}
(\qq) \> D^{\,}_{1 \la^{\,}_X; \la^{\,}_q \la^{\,}_{\bar q}} \>
D^{*}_{-1 \la^{\,}_X; \la^\prime_q \la^\prime_{\bar q}}
\simeq \rho_{+-;-+}(\qq) \> D^{\,}_{10;+-} \> D^{*}_{-10;-+}
\label{las+-}
\eeq
and [see Eqs. (\ref{polfra}) and (\ref{usfra})]
\beq
D_q^V \simeq \sum_X (1+ \alpha_q^V) \, \vert D_{10;+-} \vert^2 \,.
\label{dqv}
\eeq
In obtaining Eq. (\ref{dqv}), following Ref. \cite{abmq}, we have neglected
quark masses, we have assumed that $\pm$ helicity quarks do not fragment into
$\mp 1$ helicity vector mesons and we have defined, for each quark flavour:
\beq
\vert D_{0-1;+-} \vert^2 = \alpha^V \> \vert D_{10;+-} \vert^2 \,,
\label{alp}
\eeq
with $\alpha^V$ measured by
\beq
\alpha^V = {\rho^{\,}_{0,0}(V) \over 1 - \rho^{\,}_{0,0}(V)} \,\cdot
\label{alrho}
\eeq

We then obtain, exploiting parity invariance in quark fragmentation and
with the same assumptions as in Ref. \cite{abmq}:
\barr
\rho^{\,}_{1,-1}(V) &\>& \!\!\!\!\!\!\!
\frac{d^4\sigma^{NN \to V X}} {dx_1 \> dx_2 \> d\cos\theta^* \> dx_{_E}} 
\simeq \label{nnv1-1} \\
&& \sum_X g(x_1) \> g(x_2) \> \frac{d\hat\sigma^{gg \to \qq}}{d\cos\theta^*}
\> \rho^{gg \to \qq}_{+-;-+}(\qq;\theta^*) \> \vert D^{\,}_{10;+-}(x_{E}) 
\vert^2 \,, \nonumber
\earr
with
\beq
\frac{d^4\sigma^{NN \to VX}} {dx_1 \> dx_2 \> d\cos\theta^* \> dx_{_E}} \simeq
\sum_X  g(x_1) \> g(x_2) \>
\frac{d\hat\sigma^{gg \to \qq}}{d\cos\theta^*} \> (1 + \alpha_q^V) \>
\vert D^{\,}_{10;+-}(x_{_E}) \vert^2 \,. \label{nnvcs}
\eeq
$\theta^*$ is the production angle of $q$ and $V$ in the $gg$ {\it c.m.}
frame.

We then end up with the simple prediction, analogous to (\ref{old}):
\beq
\rho_{1,-1}(V) \simeq [1 - \rho^{\,}_{0,0}(V)] \>
\rho^{gg \to \qq}_{+-;-+}(\qq;\theta^*) \,.
\label{nnnew}
\eeq

\noindent
$\rho^{gg \to \qq}_{+-;-+}(\qq;\theta^*)$ can be easily computed and
 is given, neglecting quark masses, by:
\beq
\rho^{gg \to \qq}_{+-;-+}(\qq;\theta^*) = {1 \over 2} \>
{\sin^2\theta^* \over 1 + \cos^2 \theta^*} \,\cdot
\label{rhogg}
\eeq
A similar result holds when taking quark masses into account and it 
would not change our following conclusions.

Eqs. (\ref{nnnew}) and (\ref{rhogg}) hold for vector mesons inclusively produced
in $NN$ interactions, selecting events in which: two large energy and large
$p_T$ current jets are observed, the vector meson carries a large fraction
of the parent jet momentum with a small transverse component $k_\perp$, and
the dominant hard interaction is $gg \to \qq$.

The main point to be observed is that both Eq. (\ref{old}) and (\ref{nnnew})
have the same structure, but they differ in the dynamics of the
elementary processes, given respectively by $\rho^{Z}_{+-;-+}(\qq)$,
Eq. (\ref{rhozap}), and $\rho^{gg \to \qq}_{+-;-+}(\qq)$, Eq. (\ref{rhogg}).
Whereas the angular dependence is the same in both cases, the cofficients
in front are negative for $\rho^{Z}_{+-;-+}$ \cite{abmq} and positive for
$\rho^{gg \to \qq}_{+-;-+}$, making a clear testable difference;
while the values of $\rho_{1,-1}(V)$ are negative at LEP, they should
be positive in $NN$ interactions. Notice that the value of
$\rho^{gg \to \qq}_{+-;-+}(\qq)$ is the same that one obtains for
$e^-e^+ \to \qq$ processes at low energies, where the weak effects are
negligible and one only takes into account the electromagnetic contribution
\cite{abmq}.

\vskip 12pt
\goodbreak
\noindent
{\bf 3 -} {\mbox{\boldmath $\rho^{\,}_{1,-1}(V)$}} {\bf in the process}
{\mbox{\boldmath $\gamma N \to V + X$}}
\nobreak
\vskip 12pt
There are two classes of contributions for this process. In the first
class, usually called the resolved photon case, the photon interacts
through its partonic components; the reaction may thus be considered as a
particular case of the previous one,
$AB \to VX$, the gluon content of the photon replacing the gluon content
of the proton; the second class is defined as a direct interaction
of the photon with the partons from the nucleon and the subprocess
leading to a heavy meson production is then $\gamma g \to \qq$; under the
same assumptions and kinematical conditions we have, analogously to Eqs.
(\ref{nnv1-1}) and (\ref{nnvcs}):
\barr
\rho^{\,}_{1,-1}(V) &\>& \!\!\!\!\!\!\!\!\!
\frac{d^3\sigma^{\gamma N \to V X}} {dx \> d\cos\theta^* \> dx_{_E}} \simeq
\label{gnv1-1} \\
&& \sum_X g(x) \> \frac{d\hat\sigma^{\gamma g \to \qq}}{d\cos\theta^*}
\> \rho^{\gamma g \to \qq}_{+-;-+}(\qq;\theta^*) \>
\vert D^{\,}_{10;+-}(x_{_E}) \vert^2
\,, \nonumber
\earr
with
\beq
\frac{d^3\sigma^{\gamma N \to V X}} {dx \> d\cos\theta^* \> dx_{_E}} \simeq
\sum_X g(x) \> \frac{d\hat\sigma^{\gamma g \to \qq}}{d\cos\theta^*} \>
(1 + \alpha_q^V) \> \vert D^{\,}_{10;+-}(x_{_E}) \vert^2 \,. \label{gnvcs}
\eeq
where $\theta^*$ is the production angle of $q$ and $V$ in the $\gamma g$
{\it c.m.} frame.

Again, we end up with:
\beq
\rho_{1,-1}(V) \simeq [1 - \rho^{\,}_{0,0}(V)] \>
\rho^{\gamma g \to \qq}_{+-;-+}(\qq;\theta^*) \,,
\label{gnnew}
\eeq
where, neglecting quark masses:
\beq
\rho^{\gamma g \to \qq}_{+-;-+}(\qq;\theta^*) = {1 \over 2} \>
{\sin^2\theta^* \over 1 + \cos^2 \theta^*} \,\cdot
\label{rhopg}
\eeq
Notice that the value of $\rho^{\gamma g \to \qq}_{+-;-+}(\qq)$ is the same
as $\rho^{gg \to \qq}_{+-;-+}(\qq)$: no trace of colour dynamics remains
in the latter. This enables to add the two classes of contributions
for $\gamma N$ interactions and get a simple clear conclusion:

\noindent
Eqs. (\ref{gnnew}) and (\ref{rhopg}) being the same as (\ref{nnnew}) and
(\ref{rhogg}), the value of $\rho_{1,-1}(V)$ for $\phi$,
$D^*$ or $B^*$ mesons with large $x_{_E}$ and small $k_\perp$ inside a current
jet, should be the same in $NN$ and $\gamma N$ interactions.

\vskip 12pt
\goodbreak
\noindent
{\bf 4 -} {\mbox{\boldmath $\rho^{\,}_{1,-1}(V)$}} {\bf in the process}
{\mbox{\boldmath $\ell N \to \ell + V + X$}}
\nobreak
\vskip 12pt

Such a process can be seen as a $\gamma^* N \to VX$ process;
then one obtains, similarly to previous cases:
\beq
\rho_{1,-1}(V) \simeq [1 - \rho^{\,}_{0,0}(V)] \>
\rho^{\gamma^* g \to \qq}_{+-;-+}(\qq) \,.
\label{gsnnew}
\eeq

The difference with the previous two cases is in the value of
$\rho^{\gamma^* g \to \qq}_{+-;-+}(\qq)$. Whereas in $NN$ and $\gamma N$
the initial $\gamma$ and $g$ were unpolarized, now the off-shell photons
emitted by the lepton are polarized and one has:
\beq
\rho^{\,}_{\la^{\,}_q \la^{\,}_{\bar q}; \la^\prime_q \la^\prime_{\bar q}}
(\qq) = {1 \over N_q} \sum_{\la^{\,}_\gamma, \la^{\prime}_\gamma, \la^{\,}_g}
M^{\,}_{\la^{\,}_q \la^{\,}_{\bar q}; \la^{\,}_\gamma \la^{\,}_g} \>
M^{*}_{\la^{\prime}_q \la^{\prime}_{\bar q}; \la^{\prime}_\gamma \la^{\,}_g}
\> \rho^{\,}_{\la^{\,}_\gamma \la^{\prime}_\gamma}(\gamma^*)
\label{rhodis}
\eeq
where $\rho(\gamma^*)$ is the helicity density matrix of $\gamma^*$,
which depends on the dynamics of its emission and the DIS kinematical
variables \cite{rhogs}:
\beq
\rho^{\,}_{\la^{\,}_\gamma \la^{\prime}_\gamma}(\gamma^*_{DIS}) =
{1 \over N_\gamma^*} (-1)^{\la^{\,}_\gamma + \la^{\prime}_\gamma}
\left[ 4 l\cdot \epsilon^*(\la^{\,}_\gamma) \> 
l\cdot \epsilon(\la^{\prime}_\gamma) - Q^2 \epsilon^*(\la^{\,}_\gamma)
\cdot \epsilon(\la^{\prime}_\gamma) \right]
\label{rhostar1}
\eeq
where $l$ is the initial lepton four-momentum and $\epsilon(\la^{\,}_\gamma)$
the photon polarization vector.

By computing Eq. (\ref{rhostar1}) in the $\gamma^* g$ {\it c.m.} frame
one has:
\beq
\rho(\gamma^*_{DIS}) = {1 \over 2(2-y)^2} \left( 
\begin{array}{lll} 1 + (1-y)^2 & i C(y) e^{-i\beta} & -2(1-y) e^{-2i\beta}  \\  
-i C(y) e^{i\beta} & 4(1-y) & i C(y) e^{-i\beta} \\
-2(1-y) e^{2i\beta} & -i C(y) e^{i\beta} & 1 + (1-y)^2 
\end{array} \right)
\label{rhostar2}
\eeq
where $y=Q^2/(sx)$, $x$ is the Bjorken variable and $C(y) = 
\sqrt{2(1-y)}\,(2-y)$. $\beta$ is the azymuthal 
angle of $l$, having chosen the direction of $\gamma^*$ as the positive
$z$-axis. We choose the $\gamma^* g \to q\bar q$ scattering plane as
the $xz$ plane. 

By inserting the above expression of $\rho(\gamma^*)$ into Eq. (\ref{rhodis})
one obtains, in the $\gamma^* g$ {\it c.m.} frame and neglecting
quark masses:
\barr
\rho^{\gamma^* g \to \qq}_{+-;-+}(\qq;y,z,\theta^*,\beta) &=& 
{1 \over N_q} \Biggl\{
[(1 - z)^2 + z^2] \> [1 + (1-y)^2] + 4z(1 - z) \> (1-y) \nonumber \\
&\times& \Biggl[ -2 + {1 + \cos^2\theta^* \over 
\sin^2\theta^*} \, \cos 2\beta - 2i {\cos\theta^* \over \sin^2\theta^*} \, 
\sin 2\beta \Biggr] \label{rho+-gs} \\
&-& 4\sqrt{z(1 - z)} \> (1 - 2z) \> {C(y) \over \sqrt 2} \>
\Biggl[ {\cos\theta^* \over \sin\theta^*} \, \cos\beta - {i \over \sin\theta^*}
\, \sin\beta \Biggr] \Biggr\} \nonumber
\earr
with
\barr
N_q &=& \Biggl\{ 2 \, {1+\cos^2\theta^* \over \sin^2\theta^*} \> 
[(1 - z)^2 + z^2] \> [1 + (1-y)^2] \nonumber \\
&+& 8z(1 - z) \> (1-y) \> (2 + \cos 2\beta) \label{ngs} \\
&+& 8 \, {\cos\theta^* \over \sin\theta^*} \> \sqrt{z(1 - z)} \> (1 - 2z)
\> {C(y) \over \sqrt 2} \> \cos\beta \Biggr\} \nonumber
\earr
where $\theta^*$ is the production angle of $q$ and $V$ in the $\gamma^* g$
{\it c.m.} frame, $z=x/x_g$ and $x_g$ is the nucleon momentum fraction carried 
by the gluon. Notice that in the $Q^2 \to 0$, $x \to 0$ limit one recovers 
Eq. (\ref{rhopg}).
    
Although the dependence of Eqs. (\ref{rho+-gs}) and (\ref{ngs}) on $\beta$ 
is interesting, such angle might be difficult to measure; averaging over 
$\beta$ yields
\beq
\rho^{\gamma^* g \to \qq}_{+-;-+}(\qq;y,z,\theta^*) =
{\sin^2 \theta^* \over 2(1 + \cos^2 \theta^*)} \> 
{1 - A(y,z) \over 1 + A(y,z) \sin^2\theta^*/(1+\cos^2\theta^*)}
\label{ayz}
\eeq
where
\beq
A(y,z) = {8z(1-z)(1-y) \over [(1 - z)^2 + z^2] \> [1 + (1-y)^2]} \,\cdot
\label{defa}
\eeq

Insertion of Eqs. (\ref{ayz}) and (\ref{defa}) into Eq. (\ref{gsnnew})
gives our prediction for $\rho_{1,-1}(V)$ in DIS, and its dependence on the 
variables $y = Q^2/(sx)$, $z = x/x_g$, the production angle $\theta^*$ of $V$ 
in the $\gamma^* g$ $c.m.$ frame and the measured value of $\rho_{0,0}(V)$.

Such dependence can easily be tested experimentally. In Fig. 1 we plot
the value of $\rho^{\gamma^* g \to \qq}_{+-;-+}(\qq)$ as given by Eqs. 
(\ref{ayz}) and (\ref{defa}), as function of $\theta^*$. We have fixed
typical HERA values for the DIS variables, $Q^2 = 100$ (GeV/$c)^2$
and $x=0.01$, which imply $y=0.11$; the solid line corresponds to 
$z=0.5$, the dashed line to $z=0.02$ and the dotted line shows for
comparison $\rho^{\gamma g \to \qq}_{+-;-+}(\qq)$, Eq. (\ref{rhopg}), which 
corresponds to $z=0$. In the first case -- which means $x_g = 0.02$ -- 
the value of $\rho^{\gamma^* g \to \qq}_{+-;-+}(\qq)$ is negative, whereas
in second case -- $x_g = 0.5$ -- is positive and close to the real $\gamma$ 
value. Thus, there is a huge variation, both in size and sign, of the
value of $\rho_{1,-1}$, which should be easily observable.

The same results hold for different values of $x$ and $Q^2$, provided the
values of $y$ and $z$ do not change:  different ranges of $y$ and $z$
are reachable at different accelerators. In general, at fixed values of $y$,
the value of $\rho^{\gamma^* g \to \qq}_{+-;-+}(\qq)$ differs mostly from
the real $\gamma$ case when $z=0.5$ and one recovers Eq. (\ref{rhopg}) when
$z \to 1$ or $z \to 0$; at fixed values of $z$, instead, one reaches the
real $\gamma$ limit when $y \to 1$.

\vskip 12pt
\goodbreak
\noindent
{\bf 5 -} {\bf Conclusions}
\nobreak
\vskip 12pt

In this paper, we have derived the values of the off-diagonal element
$\rho_{1,-1}$ of the helicity density matrix of heavy vector mesons, with 
well defined properties, in jets produced through a selected hard 
interaction. Such values are crucially related to the nature of the 
partonic hard interactions and, as all quantities which depend on the 
product of different scattering amplitudes, they constitute a most 
sensitive test of the underlying dynamics. Our numerical results show
clear and unmistakable differences among the values of $\rho_{1,-1}$ of 
the same vector mesons obtained in different processes, like $e^+e^-$ 
annihilations, nucleon-nucleon and $\gamma$-nucleon processes and DIS.

Apart from the necessary approximations (like the dominance of definite 
subprocesses and the absence of any significant higher twist contributions) 
which may be controlled in the production cross section, the basic hypothesis 
underlying the whole approach is the factorizability of the hadronization 
process -- even in its detailed off-diagonal helicity structure -- from the 
hard subprocess amplitudes. In particular, the confirmation of our predictions, 
as given in Eqs. (16), (20) and (22), tests the absence of significant final 
state interactions between the jet containing the heavy vector meson and other
jets such as for instance the backward jet containing the remnants of the
target proton. The very fact of finding a non zero value of $\rho_{1,-1}$,
independently of its actual value which depends on the underlying
dynamics, shows in a clear way what the hadronization mechanism should be,
namely -- in our cases -- a $\qq$ interaction. This should be the case
at least in $NN$ interactions, as the $q \bar q$ produced in the $gg$ fusion
could be in a colour singlet state. Any experimental departure
from the results obtained here  -- no significant deviation from zero of
$\rho_{1,-1}(V)$, even for events strictly selected in order to obey our
validity conditions -- would thus constitute a minor but real crisis towards
a simple understanding of hadronization.

The measurements we suggest, and for which we give predictions, are simple 
and clear; they should help in understanding some non perturbative aspects
of strong interactions and their relation with perturbative ones.
Such an issue is of great importance and we look forward to an experimental 
study of this question, which might well be pursued with existing data.

\vskip 24pt
\noindent
{\bf Acknowledgements}
\vskip 6pt
M. A. would like to thank the Ecole Polytechnique, where this work
was initiated, and M. B. would like to thank the Department of 
Theoretical Physics of the Universit\`a di Torino, for the kind hospitality.

\vskip 24pt

\vspace{24pt}

\noindent {\Large \bf Figure Caption}

\vspace{12pt}

\noindent {\large\bf Fig. 1 :}
$\rho^{\gamma^* g \to \qq}_{+-;-+}(\qq)$ as given by Eqs. (\ref{ayz}) and 
(\ref{defa}), as function of $\theta^*$ for $Q^2 = 100$ (GeV/$c)^2$
and $x=0.01$; the solid line corresponds to $z=0.5$, the dashed line to 
$z=0.02$ and the dotted line shows for comparison 
$\rho^{\gamma g \to \qq}_{+-;-+}(\qq)$, Eq. (\ref{rhopg}), which
corresponds to $z=0$.

\clearpage

\begin{figure}
\centerline{
\epsfig{figure=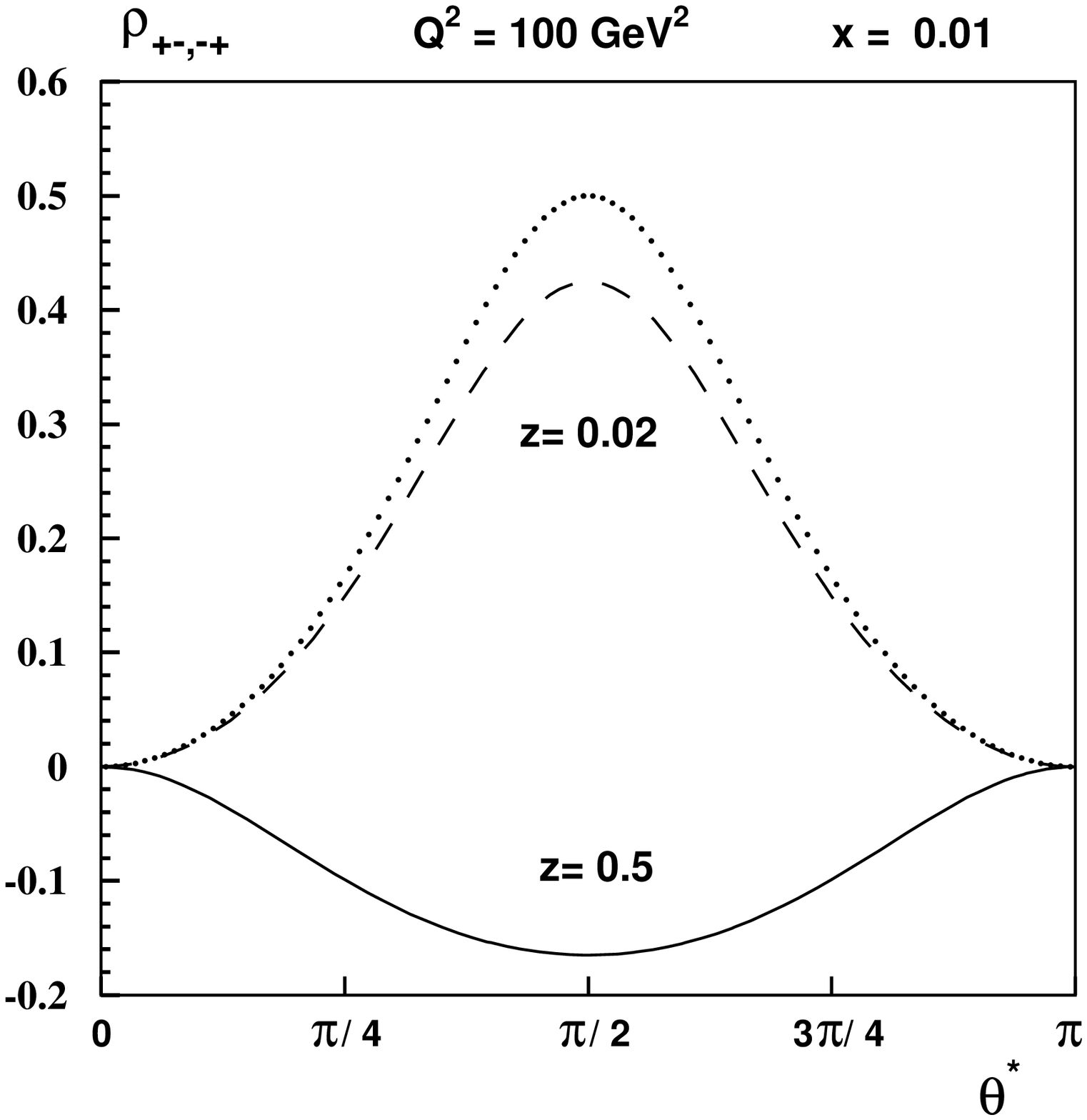,bbllx=10pt,bblly=150pt,bburx=550pt,%
bbury=680pt,width=14cm,height=14cm}}
\end{figure}

\vspace{40pt}

\begin{center}
\Huge\bf Fig. 1
\end{center}

\end{document}